\documentclass[12pt]{iopart}
\begin{document}

\title{Focus on Atomtronics-enabled Quantum Technologies}

\author{Luigi Amico $^{1,2}$ 
Gerhard Birkl $^3$, 
Malcolm Boshier $^4$, and
Leong-Chuan Kwek, $^{2,5}$ }

\address{$^1$ Dipartimento di Fisica e Astronomia, Universit\`{a} di Catania \&  CNR-MATIS-IMM \& INFN-Laboratori Nazionali del Sud, INFN, Catania  \\ 
$^2$ Centre for Quantum Technologies, National University of Singapore \\
$^3$ Technische Universit\"{a}t Darmstadt \\
$^4$ Los Alamos National Laboratories \\
$^5$ Institute of Advanced Studies and National Institute of Education, Nanyang Technological University, Singapore } 
\vspace{10pt}
\begin{indented}
\item[]February 2014
\end{indented}

\begin{abstract}
 Atomtronics is an emerging field in quantum technology that promises to realize  'atomic circuit' architectures exploiting ultra-cold atoms manipulated in versatile micro-optical circuits generated by laser fields of different shapes and intensities or micro-magnetic circuits known as atom chips. Although devising new applications for computation and information transfer is a defining goal of the field, Atomtronics  wants  to enlarge the scope of quantum simulators and to access new physical regimes with  novel fundamental science.
With  this focus issue we want to  survey the state of the art of Atomtronics-enabled Quantum Technology. We collect articles on both conceptual and applicative aspects of the field for diverse exploitations, both to extend the scope of the existing atom-based quantum devices and to devise platforms for new routes to quantum technology.
\end{abstract}

%
%
%
%
%

\section{Introduction}

The pervasive importance of electronic devices, e.g. transistors, diodes, capacitors, inductors, integrated circuits, can be observed everywhere in our daily applications and usage.  The key players in electronic devices are electrons and holes. 
Imagine circuitry in which the  carriers are atoms instead of electrons and holes. The most evident features that result from such a design would be a reduced decoherence rate due to charge neutrality of the atomic currents, an ability to realize quantum devices with fermionic or bosonic carriers, and a tunable carrier - carrier interaction from weak-to-strong, from short-to-long range, from attractive-to-repulsive in type.

The rapid progress in quantum technology is spurring this dream to reality: Atomtronics is an emerging field in physics that promises to realize those atomic circuit architectures exploiting ultra-cold atoms manipulated in versatile micro-optical circuits generated by laser fields of different shapes and intensities or by micro-magnetic circuits known as atom chips\cite{seaman2007atomtronics,pepino2009atomtronic,amico2005quantum,Amico_Atomtronics}.

With the added value of a dissipation-less flowing atomic current, Atomtronics is expected to enhance the flexibility and the scope of cold-atom quantum technology. 
With the current know-how in the field, circuits with a lithographic precision can be realized. In principle, all aspects of mesoscopic physics and devices can also be explored. It is not just classical electronics which is targeted, but also atom-based spintronics and quantum electronic structures like Josephson-junction-based circuits (SQUID devices, etc.), quantum point contacts and impurity problems. Core devices for applications in quantum metrology, e.g., nano-scale amplifiers or precision sensors, can be designed and exploited. Finally, Atomtronics may also provide new solutions for the physical realization of quantum gates for quantum information protocols and hybrid quantum systems.  

The goal of this focus issue in New Journal of Physics is to survey the state of the art of Atomtronics-enabled Quantum Technology and, at the same time, to trigger new concepts and applications for the field. We covered both theoretical and experimental works. Like many collection of works, however, it is not meant to be comprehensive nor exhaustive. Readers are encouraged to look into the relevant literature cited in these papers for a more complete understanding.

In this introduction to the focus issue we group the articles in three sections: Atomtronics Quantum Simulators, Elements for Atomtronics Integrated Circuits, and  Atomtronics Device for Sensing  and Computation.

\section{Atomtronics quantum simulators}

Quantum emulators or simulators have been one of the key goals of quantum technology\cite{dowling2003quantum,bloch2005ultracold,buluta2009quantum,
cirac2012goals,johnson2014quantum}. The Atomtronics quantum technology of atomic currents flowing in closed architectures combined with carrier statistics that can be changed easily will ultimately enable a new platform of cold-atom quantum simulators. The object of the quantum simulation may be a notoriously hard problem in physics to be solved with classical logic, or it may provide insights on problems in different areas of physics, notably  in quantum material science or high energy physics, that are still open.
Atomtronics  can address key issues,  from  fundamental concept of low-dimensional superfluidity in interacting quantum gases to the interplay of geometrical constraints, interactions and gauge potentials, to frustration effects, topological order, and edge state formation. Some examples have been addressed in recent studies in cold atomic samples , but here we are in the position to work with tunable boundary conditions, and capture physical conditions where flowing currents provide a way to actually characterize the system. The situation is comparable to developments in solid state physics: a fruitful avenue in that field is to study the (electronic) current in the system as a response to an external perturbation. The Atomtronics approach here is top-down:  Analyse the response of a quantum fluid to an external (artificial) gauge field to gain insight into the microscopic quantum dynamics. Finally, to achieve these goals, new ways to read out the state of the cold atom system  are fostered. 

\begin{itemize}
\item
{\it  Coherent superposition of current flows in an atomtronic quantum interference device}, D.Aghamalyan, M. Cominotti, M. Rizzi, D. Rossini, F. Hekking, A. Minguzzi, L.-C. Kwek and L. Amico \cite{aghamalyan2015coherent} https://arxiv.org/abs/1411.4812.

{\bf Abstract.}  We consider a correlated Bose gas tightly confined into a ring shaped lattice, in the presence of an artificial gauge potential inducing a persistent current through it. A weak link painted on the ring acts as a source of coherent back-scattering for the propagating gas, interfering with the forward scattered current. This system defines an atomic counterpart of the rf-SQUID: the atomtronics quantum interference device. The goal of the present study is to corroborate the emergence of an effective two-level system in such a setup and to assess its quality, in terms of its inner resolution and its separation from the rest of the many-body spectrum, across the different physical regimes. In order to achieve this aim, we examine the dependence of the qubit energy gap on the bosonic density, the interaction strength, and the barrier depth, and we show how the superposition between current states appears in the momentum distribution (time-of-flight) images. A mesoscopic ring lattice with intermediate-to-strong interactions and weak barrier depth is found to be a favorable candidate for setting up, manipulating and probing a qubit in the next generation of atomic experiments.
\item
{\it Entanglement and violation of classical inequalities in the Hawking radiation of flowing atom condensates},  J R M de Nova, F Sols and I Zapata \cite{de2015entanglement} https://arxiv.org/abs/1509.02224.
\\
{\bf Abstract.} We consider a sonic black-hole scenario where an atom condensate flows through a subsonicÐsupersonic interface. We discuss several criteria that reveal the existence of non-classical correlations resulting from the quantum character of the spontaneous Hawking radiation (HR). We unify previous general work as applied to HR analogs. We investigate the measurability of the various indicators and conclude that, within a class of detection schemes, only the violation of quadratic CauchyÐSchwarz inequalities can be discerned. We show numerical results that further support the viability of measuring deep quantum correlations in concrete scenarios.
\item
{\it Quantum simulation of conductivity plateaux and fractional quantum Hall effect using ultracold atoms}, N. Barber\`an, D. Dagnino, M. A. Garc'a-March, A. Trombettoni, J. Taron and M. Lewenstein\cite{barberan2015quantum}https://arxiv.org/abs/1506.06407.
\\
{\bf Abstract.} We analyze the role of impurities in the fractional quantum Hall effect using a highly controllable system of ultracold atoms. We investigate the mechanism responsible for the formation of plateaux in the resistivity/conductivity as a function of the applied magnetic field in the lowest Landau level regime. To this aim, we consider an impurity immersed in a small cloud of an ultracold quantum Bose gas subjected to an artificial magnetic field. We consider scenarios corresponding to experimentally realistic systems with gauge fields induced by rotation of the trapping parabolic potential. Systems of this kind are adequate to simulate quantum Hall effects in ultracold atom setups. We use exact diagonalization for few atoms and to emulate transport equations, we analyze the time evolution of the system under a periodic perturbation. We provide a theoretical proposal to detect the up-to-now elusive presence of strongly correlated states related to fractional filling factors in the context of ultracold atoms. We analyze the conditions under which these strongly correlated states are associated with the presence of the resistivity/conductivity plateaux. Our main result is the presence of a plateau in a region, where the transfer between localized and non-localized particles takes place, as a necessary condition to maintain a constant value of the resistivity/conductivity as the magnetic field increases.
\item
{\it Bandwidth-resonant Floquet states in honeycomb optical lattices}, A. Quelle, M. O. Goerbig and C. Morais Smith\cite{quelle2016bandwidth}https://arxiv.org/abs/1503.02635. 
\\
{\bf Abstract.} We investigate, within Floquet theory, topological phases in the out-of-equilibrium system that consists of fermions in a circularly shaken honeycomb optical lattice. We concentrate on the intermediate regime, in which the shaking frequency is of the same order of magnitude as the band width, such that adjacent Floquet bands start to overlap, creating a hierarchy of band inversions. It is shown that two-phonon resonances provide a topological phase that can be described within the Bernevig-Hughes-Zhang model of HgTe quantum wells. This allows for an understanding of out-of-equilibrium topological phases in terms of simple band inversions, similar to equilibrium systems.
\item
{\it Chaos and two-level dynamics of the atomtronic quantum interference device},  G. Arwas and D. Cohen \cite{arwas2016chaos}https://arxiv.org/abs/1510.04438.
\\
{\bf Abstract.} We study the atomtronic quantum interference device employing a semiclassical perspective. We consider an M site ring that is described by the BoseÐHubbard Hamiltonian. Coherent Rabi oscillations in the flow of the current are feasible, with an enhanced frequency due to chaos-assisted tunneling. We highlight the consequences of introducing a weak-link into the circuit. In the latter context we clarify the phaseÐspace considerations that are involved in setting up an effective 'systems plus bath' description in terms of Josephson-Caldeira-Leggett Hamiltonian.
\item
{\it Spin-orbit-coupled BoseÐEinstein-condensed atoms confined in annular potentials},  E. \"O Karabulut, F. Malet, A. L. Fetter, G. M. Kavoulakis and S. M. Reimann\cite{karabulut2016spin}https://arxiv.org/abs/1508.06761. 
\\
{\bf Abstract.} A spin-orbit-coupled Bose-Einstein-condensed cloud of atoms confined in an annular trapping potential shows a variety of phases that we investigate in the present study. Starting with the non-interacting problem, the homogeneous phase that is present in an untrapped system is replaced by a sinusoidal density variation in the limit of a very narrow annulus. In the case of an untrapped system there is another phase with a striped-like density distribution, and its counterpart is also found in the limit of a very narrow annulus. As the width of the annulus increases, this picture persists qualitatively. Depending on the relative strength between the inter- and the intra-components, interactions either favor the striped phase, or suppress it, in which case either a homogeneous, or a sinusoidal-like phase appears. Interactions also give rise to novel solutions with a nonzero circulation.
\item
{\it Quasi-molecular bosonic complexes-a pathway to SQUID with controlled sensitivity},  A. Safavi-Naini, B. Capogrosso-Sansone, A. Kuklov and V. Penna   \cite{safavi2016quasi}https://arxiv.org/abs/1601.02554.
\\
{\bf Abstract.} Recent experimental advances in realizing degenerate quantum dipolar gases in optical lattices and the flexibility of experimental setups in attaining various geometries offer the opportunity to explore exotic quantum many-body phases stabilized by anisotropic, long-range dipolar interaction. Moreover, the unprecedented control over the various physical properties of these systems, ranging from the quantum statistics of the particles, to the inter-particle interactions, allow one to engineer novel devices. In this paper, we consider dipolar bosons trapped in a stack of one-dimensional optical lattice layers, previously studied in (Safavi-Naini et al 2014 Phys. Rev. A 90 043604). Building on our prior results, we provide a description of the quantum phases stabilized in this system which include composite superfluids (CSFs), solids, and supercounterfluids, most of which are found to be threshold-less with respect to the dipolar interaction strength. We also demonstrate the effect of enhanced sensitivity to rotations of a SQUID-type device made of two CSF trapped in a ring-shaped optical lattice layer with weak links.
\item
{\it Robustness of discrete semifluxons in closed BoseÐHubbard chains}, A. Gallem\`i, M. Guilleumas, J. Martorell, R. Mayol, A. Polls and B. Juli\`a-D\`iaz \cite{gallemi2016robustness}https://arxiv.org/abs/1605.00458.
\\
{\bf Abstract.}
We present the properties of the ground state and low-energy excitations of BoseÐHubbard chains with a geometry that varies from open to closed and with a tunable twisted link. In the vicinity of the symmetric ¹-flux case the system behaves as an interacting gas of discrete semifluxons for finite chains and interactions in the Josephson regime. The energy spectrum of the system is studied by direct diagonalization and by solving the corresponding BogoliubovÐde Gennes equations. The atomÐatom interactions are found to enhance the presence of strongly correlated macroscopic superpositions of semifluxons.
\item
{\it Holographic optical traps for atom-based topological Kondo devices}, F. Buccheri, G. D. Bruce, A. Trombettoni, D. Cassettari, H. Babujian, V. E. Korepin and P. Sodano \cite{buccheri2015holographic}https://arxiv.org/abs/1511.06574.
\\
{\bf Abstract.}The topological Kondo (TK) model has been proposed in solid-state quantum devices as a way to realize non-Fermi liquid behaviors in a controllable setting. Another motivation behind the TK model proposal is the demand to demonstrate the quantum dynamical properties of Majorana fermions, which are at the heart of their potential use in topological quantum computation. Here we consider a junction of crossed TonksÐGirardeau gases arranged in a star-geometry (forming a Y-junction), and we perform a theoretical analysis of this system showing that it provides a physical realization of the TK model in the realm of cold atom systems. Using computer-generated holography, we experimentally implement a Y-junction suitable for atom trapping, with controllable and independent parameters. The junction and the transverse size of the atom waveguides are of the order of 5 ?m, leading to favorable estimates for the Kondo temperature and for the coupling across the junction. Since our results show that all the required theoretical and experimental ingredients are available, this provides the demonstration of an ultracold atom device that may in principle exhibit the TK effect.
\item
{\it Vortex dynamics in superfluids governed by an interacting gauge theory}, S. Butera, M. Valiente and P. \"Ohberg  \cite{butera2015vortex}https://arxiv.org/abs/1512.06791.
\\
{\bf Abstract.}
We study the dynamics of a vortex in a quasi two-dimensional Bose gas consisting of light-matter coupled atoms forming two-component pseudo spins. The gas is subject to a density dependent gauge potential, hence governed by an interacting gauge theory, which stems from a collisionally induced detuning between the incident laser frequency and the atomic energy levels. This provides a back-action between the synthetic gauge potential and the matter field. A Lagrangian approach is used to derive an expression for the force acting on a vortex in such a gas. We discuss the similarities between this force and the one predicted by Iordanskii, Lifshitz and Pitaevskii when scattering between a superfluid vortex and the thermal component is taken into account.
\end{itemize}

\section{Elements for Atomtronics integrated circuits}
One of the most important  goals of Atomtronics is to conceive  radically new types of quantum device exploiting phase coherence and persistent currents in optical circuits or rf-guides. Elementary circuit elements (ring condensates, matter-wave beam splitters, matter wave guides etc.) have been already constructed, with the next step being to assemble them in  atomtronic circuits with complex architectures.  So far, tunnel couplings have been the  suggested basic mechanism for coupling the circuit elements. Given the feature of the phase coherence in the system, such coupling relies on the Josephson effect.

\begin{itemize}
\item
{\it Integrated coherent matter wave circuits},  C. Ryu and M. G. Boshier \cite{ryu2015integrated}https://arxiv.org/abs/1410.8814.
\\
{\bf Abstract.} An integrated coherent matter wave circuit is a single device, analogous to an integrated optical circuit, in which coherent de Broglie waves are created and then launched into waveguides where they can be switched, divided, recombined, and detected as they propagate. Applications of such circuits include guided atom interferometers, atomtronic circuits, and precisely controlled delivery of atoms. Here we report experiments demonstrating integrated circuits for guided coherent matter waves. The circuit elements are created with the painted potential technique, a form of time-averaged optical dipole potential in which a rapidly moving, tightly focused laser beam exerts forces on atoms through their electric polarizability. The source of coherent matter waves is a Bose-Einstein condensate (BEC). We launch BECs into painted waveguides that guide them around bends and form switches, phase coherent beamsplitters, and closed circuits. These are the basic elements that are needed to engineer arbitrarily complex matter wave circuitry.
\item
{\it Atom transistor from the point of view of nonequilibrium dynamics},  Z. Zhang, V. Dunjko and M. Olshanii\cite{zhang2015atom}https://arxiv.org/abs/1506.02467.
\\
{\bf Abstract.}   We analyze the atom field-effect transistor scheme (Stickney et al 2007 Phys. Rev. A 75 013608) using the standard tools of quantum and classical nonequlilibrium dynamics. We first study the correspondence between the quantum and the mean-field descriptions of this system by computing, both ab initio and by using their mean-field analogs, the deviations from the Eigenstate Thermalization Hypothesis, quantum fluctuations, and the density of states. We find that, as far as the quantities that interest us, the mean-field model can serve as a semi-classical emulator of the quantum system. Then, using the mean-field model, we interpret the point of maximal output signal in our transistor as the onset of ergodicityÑthe point where the system becomes, in principle, able to attain the thermal values of the former integrals of motion, albeit not being fully thermalized yet.
\item
{\it Resonant wavepackets and shock waves in an atomtronic SQUID}, Y.-H. Wang, A Kumar, F. Jendrzejewski, R. M. Wilson, M. Edwards, S Eckel, G. K. Campbell and C. W. Clark, \cite{wang2015resonant}https://arxiv.org/abs/1510.02968.
\\
{\bf Abstract.} The fundamental dynamics of ultracold atomtronic devices are reflected in their phonon modes of excitation. We probe such a spectrum by applying a harmonically driven potential barrier to a $^{23}$ Na Bose-Einstein condensate in a ring-shaped trap. This perturbation excites phonon wavepackets. When excited resonantly, these wavepackets display a regular periodic structure. The resonant frequencies depend upon the particular configuration of the barrier, but are commensurate with the orbital frequency of a Bogoliubov sound wave traveling around the ring. Energy transfer to the condensate over many cycles of the periodic wavepacket motion causes enhanced atom loss from the trap at resonant frequencies. Solutions of the time-dependent Gross-Pitaevskii equation exhibit quantitative agreement with the experimental data. We also observe the generation of supersonic shock waves under conditions of strong excitation, and collisions of two shock wavepackets.
\item
{\it Stability and dispersion relations of three-dimensional solitary waves in trapped BoseÐEinstein condensates}, A. Mu\~oz Mateo and J. Brand\cite{mateo2015stability}https://arxiv.org/abs/1510.01465.
\\
{\bf Abstract.} We analyse the dynamical properties of three-dimensional solitary waves in cylindrically trapped Bose-Einstein condensates. Families of solitary waves bifurcate from the planar dark soliton and include the solitonic vortex, the vortex ring and more complex structures of intersecting vortex lines known collectively as Chladni solitons. The particle-like dynamics of these guided solitary waves provides potentially profitable features for their implementation in atomtronic circuits, and play a key role in the generation of metastable loop currents. Based on the time-dependent Gross-Pitaevskii equation we calculate the dispersion relations of moving solitary waves and their modes of dynamical instability. The dispersion relations reveal a complex crossing and bifurcation scenario. For stationary structures we find that for $\mu /{\hbar }{\omega }_{\perp }> 2.65$ the solitonic vortex is the only stable solitary wave. More complex Chladni solitons still have weaker instabilities than planar dark solitons and may be seen as transient structures in experiments. Fully time-dependent simulations illustrate typical decay scenarios, which may result in the generation of multiple separated solitonic vortices.
\item
{\it Transport of ultracold atoms between concentric traps via spatial adiabatic passage}, J. Polo, A.  Benseny, Th. Busch, V. Ahufinger and J. Mompart\cite{polo2016transport}https://arxiv.org/abs/1509.05627.
\\
{\bf Abstract.} Spatial adiabatic passage processes for ultracold atoms trapped in tunnel-coupled cylindrically symmetric concentric potentials are investigated. Specifically, we discuss the matter-wave analog of the rapid adiabatic passage (RAP) technique for a high fidelity and robust loading of a single atom into a harmonic ring potential from a harmonic trap, and for its transport between two concentric rings. We also consider a system of three concentric rings and investigate the transport of a single atom between the innermost and the outermost rings making use of the matter-wave analog of the stimulated Raman adiabatic passage (STIRAP) technique. We describe the RAP-like and STIRAP-like dynamics by means of a two- and a three-state model, respectively, obtaining good agreement with the numerical simulations of the corresponding two-dimensional Schršdinger equation.
\item
{\it Principles of an atomtronic transistor}, S. C. Caliga, C. J. E. Straatsma, A. A. Zozulya and D. Z. Anderson\cite{caliga2016transport}https://arxiv.org/abs/1512.04422.
\\
{\bf Abstract.}
A semiclassical formalism is used to investigate the transistor-like behavior of ultracold atoms in a triple-well potential. Atom current flows from the source well, held at fixed chemical potential and temperature, into an empty drain well. In steady-state, the gate well located between the source and drain is shown to acquire a well-defined chemical potential and temperature, which are controlled by the relative height of the barriers separating the three wells. It is shown that the gate chemical potential can exceed that of the source and have a lower temperature. In electronics terminology, the sourceÐgate junction can be reverse-biased. As a result, the device exhibits regimes of negative resistance and transresistance, indicating the presence of gain. Given an external current input to the gate, transistor-like behavior is characterized both in terms of the current gain, which can be greater than unity, and the power output of the device.
\item
{\it Transport dynamics of ultracold atoms in a triple-well transistor-like potential}, Seth C. Caliga, Cameron J. E. Straatsma and D. Z. Anderson \cite{caliga2016principles}https://arxiv.org/abs/1601.05008.
\\
{\bf Abstract.}
The transport of atoms is experimentally studied in a transistor-like triple-well potential consisting of a narrow gate well surrounded by source and drain wells. Atoms are initially loaded into the source well with pre-determined temperature and chemical potential. Energetic atoms flow from the source, across the gate, and into the drain where they are removed using a resonant light beam. The manifestation of atomÐatom interactions and dissipation is evidenced by a rapid population growth in the initially vacant gate well. The transport dynamics are shown to depend strongly on a feedback parameter determined by the relative heights of the two barriers forming the gate region. For a range of feedback parameter values, experiments establish that the gate atoms develop a larger chemical potential and lower temperature than those in the source.
\item
{\it Versatile electric fields for the manipulation of ultracold NaK molecules}, M. W. Gempel, T. Hartmann, T. A. Schulze, K. K. Voges, A. Zenesini and S. Ospelkaus  \cite{gempel2016versatile}https://arxiv.org/abs/1603.08348.
\\
{\bf Abstract.}
In this paper, we present an electrode geometry for the manipulation of ultracold, rovibrational ground state NaK molecules. The electrode system allows to induce a dipole moment in trapped diatomic NaK molecules with a magnitude up to 68$\%$ of their internal dipole moment along any direction in a given two-dimensional plane. The strength, the sign and the direction of the induced dipole moment is therefore fully tunable. The maximal relative variation of the electric field over the trapping volume is below $10^{-6}$. At the desired electric field value of $10 kV$ $cm^{-1}$ this corresponds to a deviation of $0.01 V cm^{-1}$. Furthermore, the possibility to create strong electric field gradients provides the opportunity to address molecules in single layers of an optical lattice. The electrode structure is made of transparent indium tin oxide and combines large optical access for sophisticated optical dipole traps and optical lattice configurations with the possibility to create versatile electric field configurations.
\item
{\it Suppression and enhancement of decoherence in an atomic Josephson junction}, Y. Japha, S. Zhou, M. Keil, R. Folman, C. Henkel and A. Vardi \cite{japha2016suppression}https://arxiv.org/abs/1511.00173.
\\
{\bf Abstract.}
We investigate the role of interatomic interactions when a Bose gas, in a double-well potential with a finite tunneling probability (a 'Bose-Josephson junction'), is exposed to external noise. We examine the rate of decoherence of a system initially in its ground state with equal probability amplitudes in both sites. The noise may induce two kinds of effects: firstly, random shifts in the relative phase or number difference between the two wells and secondly, loss of atoms from the trap. The effects of induced phase fluctuations are mitigated by atomÐatom interactions and tunneling, such that the dephasing rate may be suppressed by half its single-atom value. Random fluctuations may also be induced in the population difference between the wells, in which case atomÐatom interactions considerably enhance the decoherence rate. A similar scenario is predicted for the case of atom loss, even if the loss rates from the two sites are equal. We find that if the initial state is number-squeezed due to interactions, then the loss process induces population fluctuations that reduce the coherence across the junction. We examine the parameters relevant for these effects in a typical atom chip device, using a simple model of the trapping potential, experimental data, and the theory of magnetic field fluctuations near metallic conductors. These results provide a framework for mapping the dynamical range of barriers engineered for specific applications and set the stage for more complex atom circuits ('atomtronics').
\item
{\it Realizing and optimizing an atomtronic SQUID}, A. Mathey and L. Mathey\cite{mathey2016realizing}https://arxiv.org/abs/1601.05431.
\\
{\bf Abstract.}
We demonstrate how a toroidal BoseÐEinstein condensate with a movable barrier can be used to realize an atomtronic SQUID. The magnitude of the barrier height, which creates the analogue of an SNS junction, is of crucial importance, as well as its ramp-up and -down protocol. For too low of a barrier, the relaxation of the system is dynamically suppressed, due to the small rate of phase slips at the barrier. For a higher barrier, the phase coherence across the barrier is suppressed due to thermal fluctuations, which are included in our Truncated Wigner approach. Furthermore, we show that the ramp-up protocol of the barrier can be improved by ramping up its height first, and its velocity after that. This protocol can be further improved by optimizing the ramp-up and ramp-down time scales, which is of direct practical relevance for on-going experimental realizations.
\end{itemize}


\section{Atomtronics Devices for sensing  and computation}

 High precision interferometric sensors  using matter waves promise a considerable gain in sensitivity compared with the existing solutions (for rotation sensing, in particular, the gain over the light based technology, can be up to $\sim 10^{10}$, for equal enclosed areas and equal particle flux.)\cite{Barrett2014875}.  Since Bose-Einstein condensates are governed by a non-linear dynamics, Atomtronics sensors can be based on solitary waves. A special type of interferometric sensor is the Atomtronic Quantum Interference Device (AQUID) the atomic counterpart of the SQUID.  The AQUID effective dynamics are those of a qubit, so the device may provide  a new physical implementation for quantum computation with reduced decoherence. For such devices, the working point is set by applying synthetic magnetic fields/physical steering of the condensate which can induce quantum phase slips. Quantum gates, however, can be also realized through atom-solidstate hybrid schemes.
\begin{itemize}
\item
{\it Coherent quantum phase slip in two-component bosonic atomtronic circuits},  A. Gallem\`i, A. Mu\~oz Mateo, R. Mayol and M. Guilleumas  \cite{gallemi2015coherent}https://arxiv.org/abs/1509.04418. 
\\
{\bf Abstract.} Coherent quantum phase slip consists in the coherent transfer of vortices in superfluids. We investigate this phenomenon in two miscible coherently coupled components of a spinor Bose gas confined in a toroidal trap. After imprinting different vortex states, i.e. states with quantized circulation, on each component, we demonstrate that during the whole dynamics the system remains in a linear superposition of two current states in spite of the nonlinearity, and can be mapped onto a linear Josephson problem. We propose this system as a good candidate for the realization of a Mooij-Harmans qubit and remark its feasibility for implementation in current experiments with $^{87}$ Rb, since we have used values for the physical parameters currently available in laboratories.
\item
{\it Minimally destructive, Doppler measurement of a quantized flow in a ring-shaped BoseÐEinstein condensate}, A. Kumar, N. Anderson, W. D. Phillips, S. Eckel, G. K. Campbell and S. Stringari \cite{kumar2016minimally}https://arxiv.org/abs/1509.04759.
\\
{\bf Abstract.}
The Doppler effect, the shift in the frequency of sound due to motion, is present in both classical gases and quantum superfluids. Here, we perform an in situ, minimally destructive measurement, of the persistent current in a ring-shaped, superfluid BoseÐEinstein condensate using the Doppler effect. Phonon modes generated in this condensate have their frequencies Doppler shifted by a persistent current. This frequency shift will cause a standing-wave phonon mode to be 'dragged' along with the persistent current. By measuring this precession, one can extract the background flow velocity. This technique will find utility in experiments where the winding number is important, such as in emerging 'atomtronic' devices.
\item
{\it Engineering dark solitary waves in ring-trap Bose-Einstein condensates}, D. Gallucci and N. P. Proukakis \cite{gallucci2016engineering}https://arxiv.org/abs/1510.07078. 
\\
{\bf Abstract.}
We demonstrate the feasibility of generation of quasi-stable counter-propagating solitonic structures in an atomic BoseÐEinstein condensate confined in a realistic toroidal geometry, and identify optimal parameter regimes for their experimental observation. Using density engineering we numerically identify distinct regimes of motion of the emerging macroscopic excitations, including both solitonic motion along the azimuthal ring direction, such that structures remain visible after multiple collisions even in the presence of thermal fluctuations, and snaking instabilities leading to the decay of the excitations into vortical structures. Our analysis, which considers both mean field effects and fluctuations, is based on the ring trap geometry of Murray et al (2013 Phys. Rev. A 88 053615).
\item
{\it Comparative simulations of Fresnel holography methods for atomic waveguides},  V. A. Henderson, P. F. Griffin, E. Riis and A. S. Arnold \cite{henderson2016comparative}https://arxiv.org/abs/1601.07422.
\\
{\bf Abstract.}
We have simulated the optical properties of micro-fabricated Fresnel zone plates (FZPs) as an alternative to spatial light modulators for producing non-trivial light potentials to trap atoms within a lensless Fresnel arrangement. We show that binary (1 bit) FZPs with wavelength (1 $\mu m$) spatial resolution consistently outperform kinoforms of spatial and phase resolution comparable to commercial SLMs in root mean square error comparisons, with FZP kinoforms demonstrating increasing improvement for complex target intensity distributions. Moreover, as sub-wavelength resolution microfabrication is possible, FZPs provide an exciting possibility for the creation of static cold-atom trapping potentials useful to atomtronics, interferometry, and the study of fundamental physics.
\item
{\it  BoseÐEinstein condensation in large time-averaged optical ring potentials}, T. A. Bell, J. A. P. Glidden, L. Humbert, M. W. J. Bromley, S. A. Haine, M. J. Davis, T. W. Neely, M. A. Baker and H. Rubinsztein-Dunlop\cite{bell2016bose}https://arxiv.org/abs/1512.05079.
\\
{\bf Abstract.}
Interferometric measurements with matter waves are established techniques for sensitive gravimetry, rotation sensing, and measurement of surface interactions, but compact interferometers will require techniques based on trapped geometries. In a step towards the realisation of matter wave interferometers in toroidal geometries, we produce a large, smooth ring trap for Bose-Einstein condensates using rapidly scanned time-averaged dipole potentials. The trap potential is smoothed by using the atom distribution as input to an optical intensity correction algorithm. Smooth rings with a diameter up to $300 \mu m$ are demonstrated. We experimentally observe and simulate the dispersion of condensed atoms in the resulting potential, with good agreement serving as an indication of trap smoothness. Under time of flight expansion we observe low energy excitations in the ring, which serves to constrain the lower frequency limit of the scanned potential technique. The resulting ring potential will have applications as a waveguide for atom interferometry and studies of superfluidity.
\item
{\it Addressed qubit manipulation in radio-frequency dressed lattices},  G. A. Sinuco-Le\`on and B. M. Garraway P \cite{sinuco2016addressed}.
\\
{\bf Abstract.}
Precise control over qubits encoded as internal states of ultracold atoms in arrays of potential wells is a key element for atomtronics applications in quantum information, quantum simulation and atomic microscopy. Here we theoretically study atoms trapped in an array of radio-frequency dressed potential wells and propose a scheme for engineering fast and high-fidelity single-qubit gates with low error due to cross-talk. In this proposal, atom trapping and qubit manipulation relies exclusively on long-wave radiation making it suitable for atom-chip technology. We demonstrate that selective qubit addressing with resonant microwaves can be programmed by controlling static and radio-frequency currents in microfabricated conductors. These results should enable studies of neutral-atom quantum computing architectures, powered by low-frequency electromagnetic fields with the benefit of simple schemes for controlling individual qubits in large ensembles.
\item
{\it An atomtronic flux qubit: a ring lattice of BoseÐEinstein condensates interrupted by three weak links}, D. Aghamalyan, N. T. Nguyen, F. Auksztol, K. S. Gan, M. Martinez Valado, P. C. Condylis, L.-C. Kwek, R. Dumke and L. Amico \cite{aghamalyan2015atomtronic}https://arxiv.org/abs/1512.08376.
\\
{\bf Abstract.}
We study a physical system consisting of a Bose-Einstein condensate confined to a ring shaped lattice potential interrupted by three weak links. The system is assumed to be driven by an effective flux piercing the ring lattice. By employing path integral techniques, we explore the effective quantum dynamics of the system in a pure quantum phase dynamics regime. Moreover, the effects of the density's quantum fluctuations are studied through exact diagonalization analysis of the spectroscopy of the Bose-Hubbard model. We demonstrate that a clear two-level system emerges by tuning the magnetic flux at degeneracy. The lattice confinement, platform for the condensate, is realized experimentally employing a spatial light modulator.
\item
{\it Matter-wave interferometers using TAAP rings}, P. Navez, S. Pandey, H. Mas, K. Poulios, T. Fernholz and W. von Klitzing\cite{navez2016matter}https://arxiv.org/abs/1604.01212.
\\
{\bf Abstract.}
We present two novel matter-wave Sagnac interferometers based on ring-shaped time-averaged adiabatic potentials, where the atoms are put into a superposition of two different spin states and manipulated independently using elliptically polarized rf-fields. In the first interferometer the atoms are accelerated by spin-state-dependent forces and then travel around the ring in a matter-wave guide. In the second one the atoms are fully trapped during the entire interferometric sequence and are moved around the ring in two spin-state-dependent 'buckets'. Corrections to the ideal Sagnac phase are investigated for both cases. We experimentally demonstrate the key atom-optical elements of the interferometer such as the independent manipulation of two different spin states in the ring-shaped potentials under identical experimental conditions.
	
\end{itemize}





\end{document}